\documentclass[dvips]{article}
\usepackage{icrctc07}

\title{Recent Progress of GaAsP HPD development for the MAGIC telescope project}
\shorttitle{GaAsP HPDs for the MAGIC telescope project}
\authors{T.Y. Saito$^1$, M. Shayduk$^{1,2}$, M.V. Fonseca,$^{3}$, M. Hayashida$^1$, E.~Lorenz$^{1,4}$, K. Mannheim$^{5}$, R. Mirzoyan$^1$, T. Schweizer$^1$, M.~Teshima$^1$\\On Behalf of the MAGIC Collaboration}
\shortauthors{T.Y Saito and et al}
\afiliations{$^1$Max-Planck-institut fuer Physik, D-80805, Muenchen, Germany\\
$^2$Humboldt-Universitaet zu Berlin, D-12489, Berlin, Germany\\
$^3$Universidad Complutense, E-28040 Madrid, Spain\\
$^4$ETH Zurich, CH-8093 Switzerland\\
$^5$Universitaet Wuerzburg, D-97074 Wuerzburg, Germany}

\email{tysaito@mppmu.mpg.de}

\abstract{Today the Hybrid Photon Detector (HPD) is one of the few low light level
(LLL) sensors that can provide an excellent single and multiple 
photoelectron (ph.e.) amplitude resolution.
 The recently developed HPDs for the MAGIC telescope project with a GaAsP photocathode, namely the R9792U-40, 
provide a peak quantum efficiency (QE) of more than 50\% and a pulse width of $\sim$2 nsec. 
In addition, the afterpulsing rate of these tubes is very low 
compared to that of conventional photomultiplier tubes (PMTs), 
i.e. the value is ~300 times lower.
Photocathode aging measurements showed 
life time of more than 10 years under standard operating conditions of the  Cherenkov
Telescopes.
 Here we want to report on the recent progress with the above mentioned HPDs.}

\begin{document}
\maketitle

\section{Introduction}
The technique of the Imaging Atmospheric Cherenkov Telescopes (IACTs) has successfully been
demonstrated as powerful tools for ground-based sub-TeV and TeV gamma-ray astronomy.
Approximately 50 sources have already been discovered by IACTs. 

However, the energy range from 10 GeV to 100 GeV has not yet been well investigated.
It is very important to lower the energy threshold of IACTs and observe more in this energy range because a lot of interesting physics remains there, e.g. high red-shift AGNs and GRBs which are not observable in the TeV range 
because of absorption by Extra-galactic Background Light (EBL), new categories of sources such as LBLs, which have a lower inverse
Compton peak than HBLs in general, and the pulsed emission from galactic pulsars.

The MAGIC (Major Atmospheric Gamma-Imaging Cherenkov) telescope\cite{magic}, with a reflector diameter of
17m, is the world's largest IACT. Since fall 2003 it has been in operation on the Canary Islands of La Palma
(28.75$^{\circ}$ N, 17.90$^{\circ}$ W and 2200 m a.s.l). In order to further lower the threshold energy and 
increase the sensitivity, a second 17-m diameter telescope, located at 85-m distance from the first
 telescope, is being constructed. We call this stereoscopic observation by two telescopes the MAGIC-II project.
In the MAGIC-II project, in addition to the gain of stereoscopic observation, we are planning to use
high quantum efficiency (QE) Hybrid PhotoDetectors (HPDs) with GaAsP photocathodes \cite{Icrc2005}\cite{gaasp2} \cite{gaasp1} as alternative photo 
sensors to PMTs, which are used in IACTs.
 
\begin{figure}[t]
\begin{center}
\includegraphics [width=0.45\textwidth]{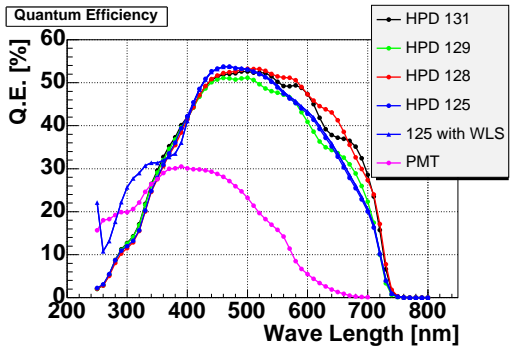}
\caption{Q.E. curves of four HPDs without WLS coating, an HPD with WLS coating and a PMT with WLS coating.}\label{QE}
\end{center}
\end{figure}

\begin{figure}[t]
\begin{center}
\includegraphics [width=0.4\textwidth]{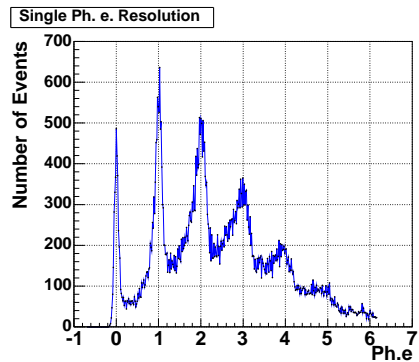}
\caption{Signal amplitude resolution with overall gain of $\sim$78000.}\label{single}
\end{center}
\end{figure}

 An HPD R9792U-40 consists of a GaAsP photocathode and of an Avalanche Diode (AD) serving as an anode. 
When applying
a $\sim$8 kV high tension to the photocathode, photoelectrons are accelerated in the high electric field and bombarded to the AD. This electron bombardment produces $\sim 1600$ electron-hole pairs per one photoelectron. These electrons subsequently induce avalanches in the active volume of AD and provide an additional
gain of 30-50 with a bias voltage of a few hundred volts, which is called avalanche amplification.
 As shown in figure \ref{QE}, Quantum Efficiency (QE) is well over 50\% at around 500 nm. Coating with WLS increases QE in the UV. One can see a large difference between HPDs and a PMT which is used in the current MAGIC camera.
Rise time, fall time and FWHM of output signals are $\sim 0.8$ nsec, $\sim 1.6$ nsec and $\sim 1.6$ nsec, respectively, when applying photocathode voltage of -8 kV and AD bias voltage of +439V.
 Multi-photoelectron peaks are well resolved as shown in figure \ref{single}.

 In addition to the favorable characteristics mentioned above, afterpulsing probability should also be cared for
in order to lower the trigger threshold. Temperature compensation of avalanche gain is also important for a stable operation. Here we will report the results of afterpulsing probability measurement and the development of a temperature compensation circuit.


\section{Afterpulsing Probability}
 
We measured afterpulsing probability by using a LED pulser (603nm) and a FADC (Acqiris cc103).
An HPD with a photocathode voltage of --8 kV (bombardment gain of 1550) and with an AD bias voltage of 370 V (avalanche gain of 30) was put in a dark box and the LED pulser illuminated it through a optical fiber. 
The LED pulser created a trigger and output signals from the HPD were recorded by the FADC with a 2 GHz sampling rate
for 500 nsec after the main pulse to search afterpulses. Two different LED light intensities were used (3 ph.e. level and 90 ph.e. level).

\begin{figure}[t]
\begin{center}
\includegraphics [width=0.4\textwidth]{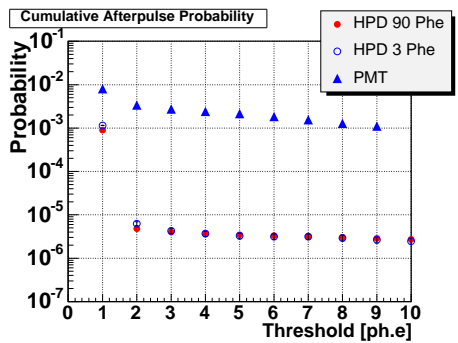}
\end{center}
\caption{Afterpulsing probability of an HPD R9792U-40 and a PMT which is used in the current MAGIC camera as a function of threshold level. Above two ph.e., the probability of the HPD is more than 300 times lower than that of the PMT. Apart from ion-feedbacks, photon-feedbacks may exist which cause afterpulses 100 times more often than ion-feedbacks but always produce 1 ph.e. afterpulses.  See text.}\label{approb}
\end{figure}

\begin{figure}[t]
\begin{center}
\includegraphics [width=0.48\textwidth]{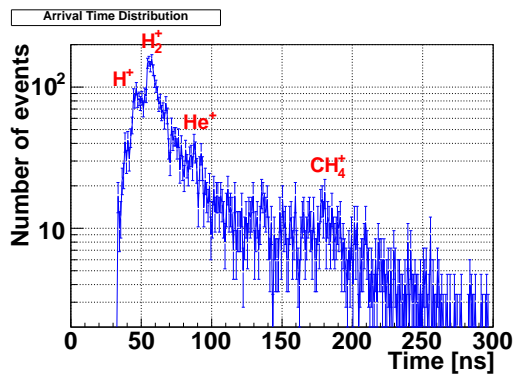}
\end{center}
\caption{Distribution of time difference between a main pulse and afterpulses of a level above 2 ph.e.. Several peaks can be seen at $\sim$45, $\sim$60, $\sim$90, $\sim$135, and $\sim$180 nsec, some of them are not clear, though. Candidate ions are written over the peaks. }\label{aptime}
\end{figure}

Figure \ref{approb} shows the results of afterpulsing probability as a function of threshold level. The afterpulsing probability ($P_{AP}$) is defined as follows,
\begin{equation}
P_{AP} = \frac{N_{AP}}{N_{MP}\times M_{MP}}
\end{equation}
where $N_{AP}$, $N_{MP}$ and $M_{MP}$ are the number of afterpulses, the number of main pulses, and the number of photoelectrons in the main pulse, respectively.
Open and red circles show HPD results with
3 ph.e. and 90 ph.e. LED light intensity, respectively, and blue triangles show the results of a PMT of the same type used in the camera of the first MAGIC telescope. The probability of HPDs to
produce afterpulses of a level above 2 ph.e. is more than 300 times less than that of the PMT. \\
 Arrival time distribution of afterpulse of a level above 2 ph.e. is shown in figure \ref{aptime}.
Several peaks can be seen ($\sim$45, $\sim$60, $\sim$90, $\sim$135, and $\sim$180 nsec), some of them are not clear, though. They can be well explained as follows. Molecules on the surface of the AD are ionized by impingement of photoelectrons with a certain probability. The ions are accelerated back and hit the photocathode, resulting in additional electron emission. 
The delay time from main pulse to afterpulse can be roughly estimated since the dimension of the HPD and the applied
voltage are known. Assuming a uniform electric field with --8 kV of potential difference and the distance of 2.8 cm between the photocathode and the AD, the delay time can be estimated as $\sim$45$\sqrt{\frac{M/M_p}{Z}}$ nsec, where $M, M_p, Z$ is the mass of ion, the mass of proton, and the charge of ion, respectively. Peaks seen in
figure \ref{aptime} may reflect feedbacks of ions with $\frac{M/M_p}{Z} = 1, 2, 4,8,16$, where protons, hydrogen molecular ions, helium ions, and methane ions can be the candidates.

If the trigger threshold of afterpulse is set at a 1 ph.e. level the afterpulsing probability increases
by two orders of magnitude. 
Apart from ion-feedbacks, there may be another mechanism, which
always produces single-ph.e.-level afterpulses.
 One possible explanation of this additional mechanism is a photon-feedback. Scintillation emission from ceramics inside the HPD is one of possible candidates for the origin
of photon-feedback. Arrival time distribution of one ph.e level afterpulse does not show any peak structure,
but an exponential decay structure, which supports the explanation above.

It is widely known that GaAsP photocathodes have a higher Q.E. but a shorter lifetime than e.g. bialkali ones. 
Cesium layer on the GaAsP crystal shows a larger degradation by ion-feedbacks. However, since the HPD R9792U-40s have a low ion-feedback rate, their lifetime is long enough for them to be used
for IACTs. Lifetimes of 5 HPDs have been measured and, if we define the lifetime as the period when Q.E. degrades by relatively 20\% and that in 1 year we have 1000 hours of operation under 300 MHz night sky background photons, all of them have more than 10 years of life.

\section{Temperature Compensation}

\begin{figure}[t]
\begin{center}
\includegraphics [width=0.4\textwidth]{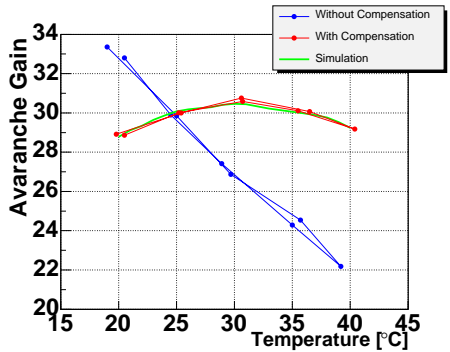}
\end{center}
\caption{Temperature dependence of the avalanche gain. The blue, red and green lines denote change of the gain without a compensation circuit ($\sim-2$\%/$^{\circ}$C), with the compensation circuit ($\sim -0.3$\%/$^{\circ}$C from 25$^{\circ}$ to 35$^{\circ}$) and simulation result of temperature compensation.}\label{temp-comp}
\end{figure}

Avalanche amplification strongly depends on the working temperature. We measured $\sim-2\%/^{\circ}$C (see figure \ref{temp-comp}) at gain of 30 at 25$^{\circ}$C. This dependence is one order of magnitude stronger than that of PMTs and should be compensated. We developed a temperature compensation circuit with
3 resistors, a DC/DC converter (APD 5P501201, Systems Development \& Solutions). and a thermistor (103AT-2, Ishizuka Electronic Corporation). As the temperature goes higher, a higher bias voltage is applied to the AD through the circuit. 
An HPD with the compensation circuit was put in a temperature regulation chamber and the LED pulser illuminated it with a light intensity of several ph.e.. Output charge distribution was recorded at different temperatures and the change of the gain was estimated by using the single ph.e. peak. In order to make sure that temperature in the chamber was well 
stabilized and there was no hysteresis, we measured it twice, i.e. first the temperature was raised from about 20$^{\circ}$C to about 40$^{\circ}$C and then lowered to about 20$^{\circ}$C.
Figure \ref{temp-comp} shows the result. Blue and red points show the temperature dependence of the avalanche gain without and with the compensation circuit, respectively. A green line denotes the simulation result.
Temperature dependence was suppressed at the level of $\sim$0.3\%/1 $^{\circ}$C from 25 $^{\circ}$C to 35$^{\circ}$C, which is the same level as that of PMTs.
It should be noted that we tuned the system for a mean temperature of 30 degrees this time, but that we can easily shift it
by changing the resistors of the circuit.


\section{Summary} 
The advantage of the HPD R9792U-40 from Hamamatsu compared to conventional PMTs is not only a higher quantum efficiency but also a 300 times lower afterpulsing probability. Extremely low afterpulsing probability leads to a long 
lifetime, estimated to be more than 10 years under standard conditions of IACTs. Temperature dependence of the avalanche gain can be reduced at the same level of a PMT by using a simple compensation circuit based on a thermistor.
The field test of HPDs are scheduled using the second telescope.

\section{Acknowledgements}
The MAGIC project is supported by the MPG (Max-Planck-Society in Germany), and the BMBF (Federal Ministry of Education and Research in Germany), the INFN (Italy), the CICYT (Spain) and the IAC (Instituto de Astrophysica de Canarias).

\nocite{ref4}
\nocite{ref5}
\nocite{ref6}
\nocite{ref7}
\bibliography{icrc2007_saito.bib}
\bibliographystyle{plain}

\end{document}